\documentclass[preprint,11pt]{aastex}

\newcommand{\BS}{BS~16934--002} 
\newcommand{\HD}{HD~122563} 
\newcommand{\kms}{~km~s$^{-1}$} 
\newcommand{\teff}{$T_{\rm eff}$}
\newcommand{\logg}{$\log g$}
\newcommand{\loggf}{$\log gf$}
\newcommand{\vt}{$v_{\rm micro}$}
\newcommand{\loge}{$\log \epsilon$}
\newcommand{\ebv}{$E(B-V)$}
\shorttitle{Abundance analysis of {\BS}}
\shortauthors{Aoki et al.}

\begin{document}

\title{Spectroscopic Studies of Extremely Metal-Poor Stars with the
Subaru High Dispersion Spectrograph. IV. The
$\alpha$-Element-Enhanced Metal-Poor Star {\BS}\footnote{Based on data
collected at the Subaru Telescope, which is operated by the National
Astronomical Observatory of Japan.}}

\author{Wako Aoki\altaffilmark{2}, Satoshi Honda\altaffilmark{2},
Timothy C. Beers\altaffilmark{3}, Masahide
Takada-Hidai\altaffilmark{4}, Nobuyuki Iwamoto \altaffilmark{5}, 
Nozomu Tominaga\altaffilmark{6}, Hideyuki Umeda\altaffilmark{6},
Ken'ichi Nomoto\altaffilmark{6}, John E. Norris\altaffilmark{7}, Sean
G. Ryan\altaffilmark{8, 9}}

\altaffiltext{2}{National Astronomical Observatory, Mitaka, Tokyo,
181-8588 Japan; email: aoki.wako@nao.ac.jp, honda@optik.mtk.nao.ac.jp}
\altaffiltext{3}{Department of Physics and Astronomy, CSCE: Center for
the Study of Cosmic Evolution, and JINA: Joint Institute for Nuclear Astrophysics,
Michigan State University, East Lansing, MI 48824-1116; email: beers@pa.msu.edu}
\altaffiltext{4}{Liberal Arts Education Center, Tokai University,
Hiratsuka, Kanagawa, 259-1292, Japan; email:
hidai@apus.rh.u-tokai.ac,jp}
\altaffiltext{5}{Japan Atomic Energy Agency, 
Ibaraki 319-1195, Japan; email:
iwamoto.nobuyuki@jaea.go.jp}
\altaffiltext{6}{Department of Astronomy, School of Science,
University of Tokyo, Bunkyo-ku, Tokyo 113-0033, Japan; email:
tominaga@astron.s.u-tokyo.ac.jp,
umeda@astron.s.u-tokyo.ac.jp,nomoto@astron.s.u-tokyo.ac.jp}
\altaffiltext{7}{Research School of Astronomy and Astrophysics, The
Australian National University, Mount Stromlo Observatory, Cotter
Road, Weston, ACT 2611, Australia; email: jen@mso.anu.edu.au}
\altaffiltext{8}{Department of Physics and Astronomy, The Open
University, Walton Hall, Milton Keynes, MK7 6AA, UK} 
\altaffiltext{9}{Present address: Centre for Astrophysics Research, STRI and School of Physics,
Astronomy and Mathematics, University of Hertfordshire, College Lane,
Hatfield AL10 9AB, United Kingdom; s.g.ryan@herts.ac.uk} 

\begin{abstract} 

A detailed elemental abundance analysis has been carried out for the
very metal-poor ([Fe/H] $= -2.7$) star {\BS}, which was identified in
our previous work as a star exhibiting large overabundances of Mg and
Sc. A comparison of the abundance pattern of this star with that of
the well-studied metal-poor star {\HD} indicates excesses of O, Na,
Mg, Al, and Sc in {\BS}. Of particular interest, no excess of C or N
is found in this object, in contrast to CS~22949--037 and
CS~29498--043, two previously known carbon-rich, extremely metal-poor
stars with excesses of the $\alpha$ elements. No established
nucleosynthesis model exists that explains the observed abundance
pattern of {\BS}. A supernova model, including mixing and fallback,
assuming severe mass loss before explosion, is discussed as a
candidate progenitor of {\BS}.

\end{abstract}


\section{Introduction}\label{sec:intro}

Observational studies of stellar chemical compositions over the past few decades
have revealed the presence of moderate overabundances in $\alpha$ elements
(e.g., O, Mg, Ca) in the majority of metal-deficient stars in the halo of the
Galaxy \citep[e.g. ][ and references therein]{mcwilliam97}. This observation is
usually interpreted as arising from the expected dominant contribution of the
yields of core-collapse supernovae in the early Galaxy, as compared to the
yields of Type-Ia supernovae, which provide significant amount of iron, but at
later times.

Recently, however, a handful of metal-poor stars have been identified
with [$\alpha$/Fe] abundance ratios that are quite different from the
general trend.\footnote{[A/B] = $\log(N_{\rm A}/N_{\rm B})-
\log(N_{\rm A}/N_{\rm B})_{\odot}$, and $\log\epsilon_{\rm A} =
\log(N_{\rm A}/N_{\rm H})+12$ for elements A and B.}  A few stars are
known to exhibit quite low [$\alpha$/Fe] ratios \citep[e.g.
][]{ivans03}, while another small set of stars is known to exhibit
significantly high ratios (e.g., [Mg/Fe]$\sim$+1.8, see McWilliam et
al. 1995; Norris et al. 2001; Depagne et al.  2002; Aoki et al. 2002,
2004).\footnote{Following the nomenclature conventions of Beers \&
Christlieb (2005), we suggest that the Alpha-Deficient Metal-Poor
stars be referred to as ADMP stars, while the Carbon-Enhanced
Metal-Poor stars with $\alpha$-element enhancements be referred to as
CEMP-$\alpha$ stars.} Although the number of these anomalous stars
remains small, their presence indicates that their nucleosynthesis
histories apparently differ from the great majority of halo stars.

Our previous study \citep{aoki05} identified a new $\alpha$-enhanced,
very metal-poor star, {\BS}, with [Mg/Fe] = +1.25. A clear difference
between this star and the previously known $\alpha$-rich, metal-poor
stars (CS~22949--037 and CS~29498--043) is that no excess of C is
found in {\BS}; we suggest that such a star be referred to as an
Alpha-Enhanced Metal-Poor (AEMP) star. In order to investigate the
astrophysical origin of the $\alpha$-enhancement in this so-far unique
object, we obtained a higher-quality spectrum of {\BS}, and perform a
more detailed elemental-abundance analysis. One important result of
this study is the measurement of O and Na abundances for {\BS}, which
are also clearly higher than the typical values found in other very
metal-poor halo stars. So as to avoid uncertainties arising from
systematic errors in the abundance analyses and evolutionary effects
on the surface chemical composition, the abundance ratios of each
element are obtained with respect to the well-studied metal-deficient
star {\HD}, which has similar atmospheric parameters to those of
{\BS}.  Details of our observations and abundance analyses are
reported in \S~\ref{sec:obs} and \S~\ref{sec:ana}, respectively. We
discuss implications of the observed excesses of the $\alpha$ elements
in this star, and offer possible interpretations in \S~\ref{sec:disc}.

\section{Observations}\label{sec:obs}

High-resolution spectroscopy was obtained for {\BS} with the High Dispersion
Spectrograph \citep[HDS; ][]{noguchi02} of the Subaru Telescope (Table~\ref{tab:obs}). A
blue spectrum was obtained in our previous study \citep{aoki05}. The new
spectrum covers the wavelength range from 4030 to 6800~{\AA}, with an absence of
data from 5350 to 5450~{\AA} due to the physical gap between the two CCDs. The
resolving power is $R=60,000$, which is slightly higher than that of the
previous observation for the blue range \citep{aoki05}. For comparison purposes,
the bright metal-poor giant {\HD}, which has quite similar atmospheric
parameters to those of {\BS}, was also observed with the same spectrograph
setup. Rapidly rotating, early-type stars were observed in each run to divide
out telluric absorption features.

Standard data reduction procedures (bias subtraction, flat-fielding,
background subtraction, extraction, and wavelength calibration) are
carried out with the IRAF echelle package\footnote{IRAF is distributed
by the National Optical Astronomy Observatories, which is operated by
the Association of Universities for Research in Astronomy, Inc. under
cooperative agreement with the National Science Foundation.} as
described by \citet{aoki05}. Telluric absorption features are
corrected using the spectra of standard stars (see \S
\ref{sec:cno}). The signal-to-noise (S/N) ratios achieved are provided
in Table~\ref{tab:obs}, as estimated from the peak counts of the
spectra at $\sim 4500$~{\AA}.

\begin{figure}
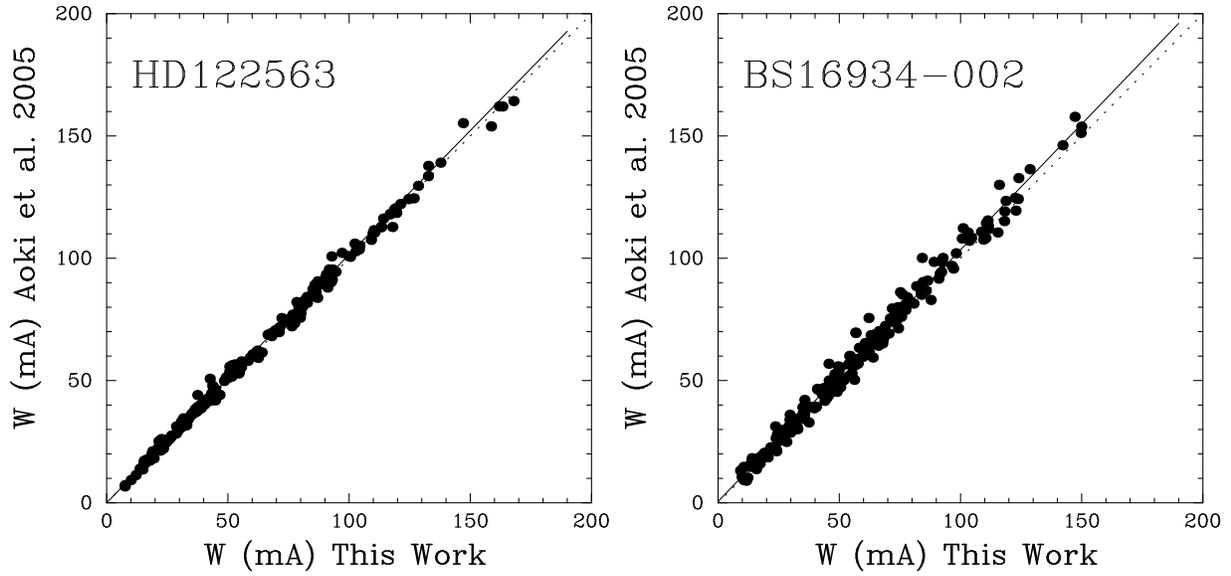
 
\includegraphics[width=8cm]{f1a.ps} 
\includegraphics[width=8cm]{f1b.ps} 
\caption[]{Comparison of the equivalent widths measured by
\citet{aoki05} with those in the present work. The solid
line indicates the result of a least-square fit, while the dotted line
shows a line with a slope of unity.
}
\label{fig:ew} 
\end{figure}

Equivalent widths for prominent absorption lines are measured by
fitting gaussian profiles. Since the spectral resolution is slightly
different between the present work and that of \citet{aoki05}, we did
not combine the spectra from these two studies. Instead, we adopt the
means of the equivalent widths for the lines in common, as measured
from the individual spectra.  Figure~\ref{fig:ew} shows comparisons of
the equivalent widths measured from these two studies. Since the
agreement is quite good, we apply no corrections to the measured
values. The equivalent widths finally adopted for the abundance
analysis are listed in Table~\ref{tab:ew}. The root-mean-square (rms)
of the difference of the two measurements is 4~m{\AA}.

Radial velocities are measured using clean \ion{Fe}{1} lines; the
results are provided in Table~\ref{tab:obs}. No significant change of
heliocentric radial velocity was found for {\BS} between the four
observations from April 2001 to May 2004 \citep[see
][]{aoki05}. Hence, there exists no evidence of binarity for {\BS}, nor
for {\HD}, based on the data obtained thus far.

The equivalent widths of the interstellar \ion{Na}{1} D1 and D2 lines for {\BS}
are 114.4 and 159.6~m{\AA}, respectively. The empirical relations between the
strength of the interstellar absorption of \ion{Na}{1}~D2 and interstellar
reddening by \citet{Munari/Zwitter:1997} provide an estimate of {\ebv} = 0.050.
This value supports the independent estimate for reddening obtained from the
dust maps of Schlegel, Finkbeiner, \& Davis (1998) ({\ebv} = 0.032), which was adopted by
\citet{aoki05} in the determination of the effective temperature for this star.

\section{Abundance Analyses and Results}\label{sec:ana}

As in our previous work, a standard analysis using the model
atmospheres of \citet{kurucz93} is performed for the measured
equivalent widths for most elements, while a spectrum synthesis
technique is applied to the molecular bands of CH and CN, and some
additional weak atomic lines. We adopt the atmospheric parameters
determined by \citet{aoki05}; these are listed in
Table~\ref{tab:param}. An exception is the microturbulent velocity
({\vt}), which is re-determined from the \ion{Fe}{1} lines by
demanding no dependence of the derived abundance on equivalent
widths. The {\vt} values agree well with the result of the previous
analysis (the difference is 0.1~km~s$^{-1}$). The abundance results
are presented in Table~\ref{tab:res}.
 
\subsection{Carbon, Nitrogen, and Oxygen}\label{sec:cno}

\begin{figure} 
\includegraphics[width=8cm]{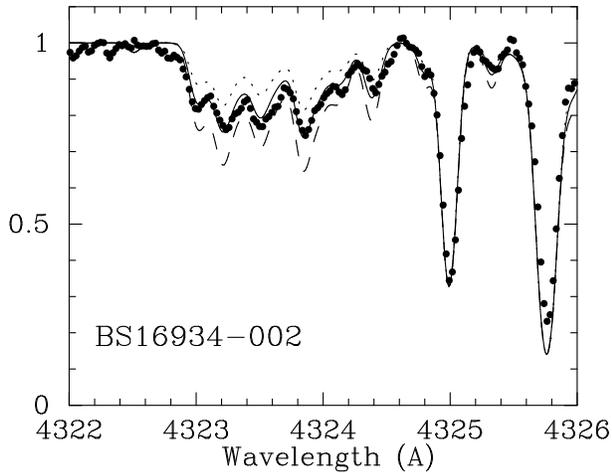} 
\caption[]{Comparison of synthetic spectra for the CH band in {\BS}
with the observed one.  The assumed abundances are [C/Fe]$= -0.27 \pm 0.20$.}
\label{fig:ch} 
\end{figure}

\begin{figure} 
\includegraphics[width=8cm]{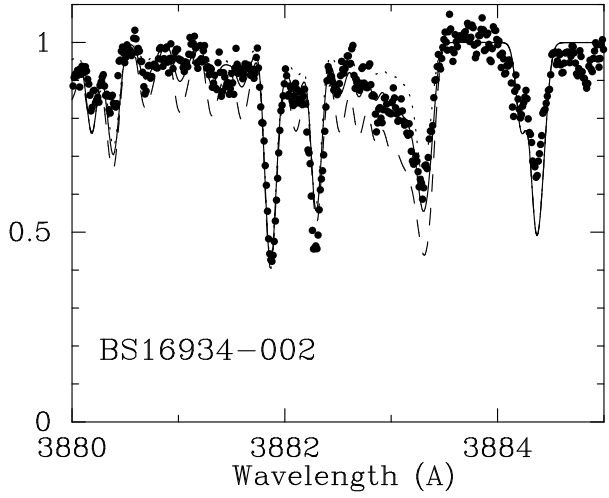} 
\caption[]{Same as Fig.~\ref{fig:ch}, but for the CN band. 
The assumed abundances are [N/Fe]$= 0.65 \pm 0.30$.}
\label{fig:cn} 
\end{figure}

The carbon abundance is re-determined from the CH 4322~{\AA} band, as
in our previous analysis, for the new spectrum of {\BS} (Figure
~\ref{fig:ch}). The agreement with the previous result is fairly good.
The nitrogen abundance is determined from the CN 3883~{\AA} band in
the blue spectrum of \citet{aoki05}, which was not analyzed in our
previous work (Figure~\ref{fig:cn}).  The absorption features of CH
and CN are quite similar for {\BS} and {\HD}. The [C/Fe] in both
objects is underabundant with respect to the solar ratio ([C/Fe]
$\simeq -0.3$), while [N/Fe] is moderately overabundant ([N/Fe]
$\simeq +0.6$, see \S \ref{sec:disc}).

\begin{figure} 
\includegraphics[width=8cm]{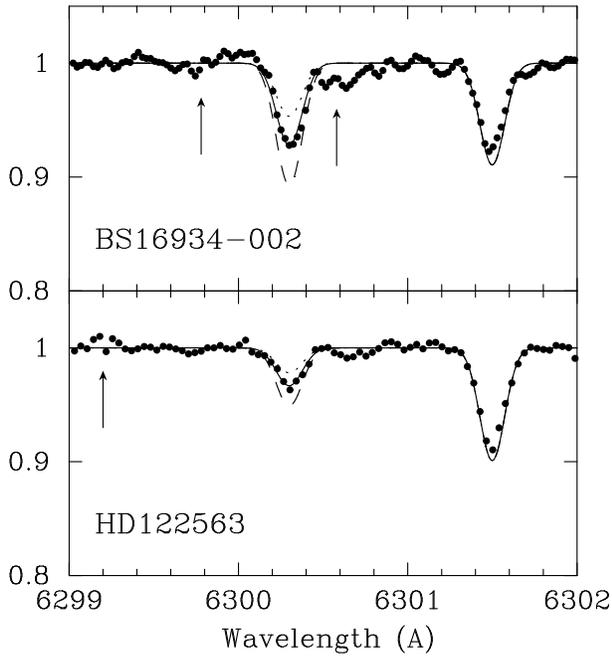} 
\caption[]{Same as Fig.~\ref{fig:ch}, but for the [\ion{O}{1}]
6300~{\AA} line for {\BS} (upper panel) and {\HD} (lower panel). The assumed
abundances are [O/Fe]$=+1.14 \pm 0.20$, and $+0.74 \pm 0.20$ for {\BS}
and {\HD}, respectively. The positions of telluric lines, which are
corrected using spectra of standard stars, are shown by the arrows.}
\label{fig:o6300} 
\end{figure}

The oxygen abundances are determined from the [\ion{O}{1}] 6300 and
6363~{\AA} lines through a standard analysis. The stellar absorption
lines of the [\ion{O}{1}] in {\BS} are well separated from sky emission lines due to
its large doppler shift. The telluric absorption lines are corrected
using the spectra of standard stars. While significant contamination
of \ion{Ni}{1}, \ion{Ca}{1}, and CN lines is reported for the solar
spectrum \citep[e.g. ][]{asplund04}, that is negligible in extremely
metal-poor giants studied here. Comparisons of synthetic spectra for
the [\ion{O}{1}] 6300~{\AA} feature with the observed spectra,
assuming three different O abundances, are shown in
Figure~\ref{fig:o6300} for {\BS} and {\HD}, where the positions of
telluric absorption features appearing in the original spectra are
indicated. A comparison of the [\ion{O}{1}] absorption in these two
stars clearly shows the excess of oxygen in {\BS} with respect to
{\HD}. We note that the abundance analysis for carbon and nitrogen are
made including the oxygen overabundance. The effect of the oxygen
excess on the derived carbon and nitrogen abundance is about 0.1~dex or
smaller.

The oxygen abundance of {\HD} was studied by \citet{barbuy03}, who used three
abundance indicators (the [\ion{O}{1}] 6300~{\AA} line, the near-infrared
\ion{O}{1} triplet, and the OH molecular lines at 1.5--1.7~$\mu$m). Our result
([O/Fe]$=+0.7$) agrees well with the value ([O/Fe]$=+0.6$) they obtained from
the [\ion{O}{1}] 6300~{\AA} line. Although there exists some discrepancy in
oxygen abundances determined from different indicators, the excess of {\BS}
estimated from the [\ion{O}{1}] feature, compared with that of {\HD}, is robust. 

\subsection{Elements from Na to Zn}\label{sec:nazn}

Standard analyses are applied to the equivalent widths listed in
Table~\ref{tab:ew}. The effects of hyperfine splitting on the Mn lines
\citep{mcwilliam95b} are included in the analysis. 

The Na abundances are determined from the weak \ion{Na}{1} lines at 5682 and
5688~{\AA}. While the \ion{Na}{1} D lines are measurable in the spectrum of
{\BS}, those in our spectrum of {\HD} severely blend with interstellar
absorption, and could not be used for abundance analysis. The Na abundance of
{\BS}, derived from the D lines, is about 0.5~dex higher than that obtained from
the two weak lines mentioned above. This discrepancy is well explained by NLTE
effects studied by \citet{takeda03}, who predicted the effect to be
approximately 0.4~dex for cool giants with an equivalent width of 200~m{\AA} for
the $\lambda$5895~{\AA} line. 

We exclude the lines of the Mg triplet at 5170~{\AA} (b lines) and at
3820~{\AA}, because these very strong lines are inappropriate for
precision abundance determinations. The Mg abundances determined from the 5528
and 5711~{\AA} lines, which are detected in our new spectrum, agree very
well with the results from the other four lines.

The Si abundances are determined from the four \ion{Si}{1} lines at
4103, 5772, 5948, and 6155~{\AA}. The very strong \ion{Si}{1}
3905~{\AA} line is excluded from the present analysis. Although the
three lines in the red region are very weak ($W_{\lambda}=3$--5~m{\AA}
in {\BS}), the agreement of the Si abundances from the individual lines
is fairly good. In contrast to the Na, Mg, and Al abundance ratios
relative to Fe, the [Si/Fe] ratio of {\BS} agrees very well with that of
{\HD}. 

While the abundance ratios of the iron-peak elements (Mn, Co, Ni) are very
similar between the two stars, significant excesses of the elements from Na to Ti,
with an exception of Si, are found in {\BS}, as compared with {\HD}. 

\subsection{Neutron-Capture Elements}\label{sec:ncap}

Both the light neutron-capture element Sr and the heavy neutron-capture element
Ba are detected in our spectra of {\BS}. In the abundance measurement for Ba the
effects of hyperfine splitting are included, using the line list of
\citet{mcwilliam98}, and assuming the isotope ratios found in the r-process
component in solar-system material \citep{arlandini99}. The heavy
neutron-capture element ratio [Ba/Fe] of {\BS}, as well as that of {\HD}, is
significantly underabundant relative to solar, as is usually found in very
metal-poor stars. This implies no significant nucleosynthesis contribution from
the s-process, although the complete abundance pattern of the heavy
neutron-capture elements is not available for {\BS}.

The [Sr/Fe] abundance ratio of {\BS} is more than 1~dex lower than
that of {\HD}. However, given the large scatter of Sr abundances found
in very metal-poor stars, the values of both stars are within the
range of the Sr abundance distribution of previous observations. It
should be noted that HD~122563 is a star that exhibits relatively
large excesses of the light neutron-capture elements \citep{honda06},
while the Sr-abundance ratio, as well as the Ba one, of {\BS} are
relatively low among stars with similar metallicity
\citep{aoki05}. This suggests that {\BS} formed from material that is
not well mixed in the early Galaxy, and might support our
interpretation that the chemical composition of this object was
basically determined by a single supernova event (\S 4).

\subsection{Uncertainties}

Random abundance errors in the standard analysis are estimated from the
standard errors of the abundances derived from individual lines for
each species. However, these values are sometimes unrealistically
small when only a few lines are detected. For this reason, we adopt
the larger of (a) the value for the listed species and (b) the
standard error derived using the standard deviation of the abundance
from individual \ion{Fe}{1} lines and the number of lines used for the
species as estimates of the random errors. Typical random errors are
on the order of 0.1~dex. 

We also tried another estimate of random errors using the standard
deviation of the difference of the two equivalent width ($W$)
measurements (4~m{\AA}) in \S2. We selected several \ion{Fe}{1} lines
around 4000~{\AA} having $W$ of 20 to 150~m{\AA}, and made abundance
calculations changing the $W$ by 4~m{\AA}. The effect on the derived
Fe abundance distributes 0.07 to 0.11~dex, depending on the $W$. This
would support the above estimate of random errors from the standard
errors of abundances derived from individual lines.

The random errors of carbon and nitrogen abundances are estimated by
the fitting uncertainties for the CH and CN absorption bands including
the uncertainties of continuum placements. We adopted 0.1~dex and
0.2~dex as these errors for carbon and nitrogen abundances. We also
re-estimate the random errors for oxygen and sodium abundances, which
are determined from weak absorption features and might be significantly
affected by the uncertainty of the continuum level. We adopt the fitting
uncertainties of 0.15 and 0.10~dex for the [\ion{O}{1}] and
\ion{Na}{1} lines, respectively. 

We adopt estimates of errors arising from uncertainties of the
atmospheric parameters from \citet{aoki05} for HD~122563, which were
evaluated for $\sigma (T_{\rm eff})=100$~K, $\sigma (\log g)=0.3$~dex,
and $\sigma (v_{\rm micro}) =0.3$~km s$^{-1}$. Finally, we derive the
total uncertainty by adding, in quadrature, the individual errors, and
list them in Table \ref{tab:res}.

\section{Discussion and Concluding Remarks}\label{sec:disc}

\subsection{Abundance Pattern of {\BS}}\label{sec:abpattern}

Abundance measurements of individual elements in metal-poor stars contain a
variety of uncertainties, in addition to random measurement errors, due to
errors in the adopted transition probabilities of spectral lines, errors in the
adopted atmospheric parameters, and from uncertain NLTE effects. In order to
avoid the influence of such uncertainties for the chemical abundances of {\BS},
we consider instead the abundance ratios (logarithmic differences) between {\BS}
and {\HD}. Although the uncertainty in the absolute abundances (e.g., [Fe/H])
remains, the difference of the abundance patterns between these two stars should
be little affected by NLTE effects and errors in the atmospheric parameters.

\begin{figure} 
\includegraphics[width=8cm]{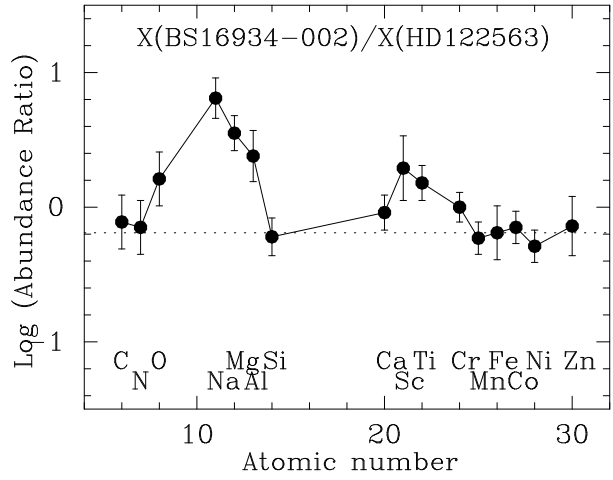} 
\caption[]{Logarithmic abundance differences between {\BS} and {\HD} as a function
of atomic number.} 
\label{fig:diff} 
\end{figure}

Figure~\ref{fig:diff} shows the difference in the abundance ratios between {\BS}
and {\HD}, on a logarithmic scale (i.e., {\loge}$_{\rm BS~16934-002}-${\loge}$_{\rm
HD~122563}$), as a function of atomic number. The abundance patterns of iron-peak
elements in these two stars exhibit excellent agreement. One exception might be
Cr -- the [Cr/Fe] value of {\BS} is 0.2~dex higher than that of {\HD}. 

By way of contrast, the abundance ratios of the elements from O to Ti
in {\BS}, exhibit significant excesses with respect to Fe. The
elements O, Na, Mg, Al, and Sc exhibit enhancements larger than
+0.4~dex, while the excesses of Ca and Ti are smaller. An exception is
Si -- the Si abundance ratio of {\BS} is almost identical to that of
{\HD}. We note that the Ti abundances derived from \ion{Ti}{1} and
\ion{Ti}{2} yield similar differences between {\BS} and {\HD},
suggesting that the difference is real. The enhancement of Na, an
element having odd atomic number, is more significant than that of
Mg. Another element with odd atomic number, Sc, also exhibits a larger
excess than those of Ca and Ti.

The $\alpha$ elements in metal-deficient stars are believed to be produced
primarily by core-collapse supernovae. The observed excesses of $\alpha$ elements in {\BS}
suggests the possible contribution of ``special'' core-collapse supernovae,
which yield high $\alpha$-to-iron abundance ratios. There are two extremely
metal-poor (EMP) CEMP-$\alpha$ stars known, CS~22949--037 and CS~29498--043, which
exhibit large enhancements of the $\alpha$ elements \citep{norris01,aoki02d,
depagne02}. \citet{umeda03,umeda05} interpreted the excesses of the $\alpha$
elements, as well as of C and N, in these objects as arising from so-called
``faint supernovae'', for which they modeled the mixing after explosive
nucleosynthesis and subsequent fallback onto the collapsed remnant
\citep[see also ][]{tsujimoto03}. However, the C abundance of {\BS} is
not enhanced, in contrast to that of CS~22949--037 and CS~29498--043. Moreover, the
total abundance of C and N in this star exhibits no significant excess ([(C+N)
/Fe] = +0.14). The normal C and N abundances, along with excesses of O and Mg,
are unique characteristics of {\BS}, the first known AEMP star.

\subsection{A Possible Interpretation of the Peculiar Composition of {\BS}}\label{sec:model}

\begin{figure} 
\begin{center}
\includegraphics[width=8cm]{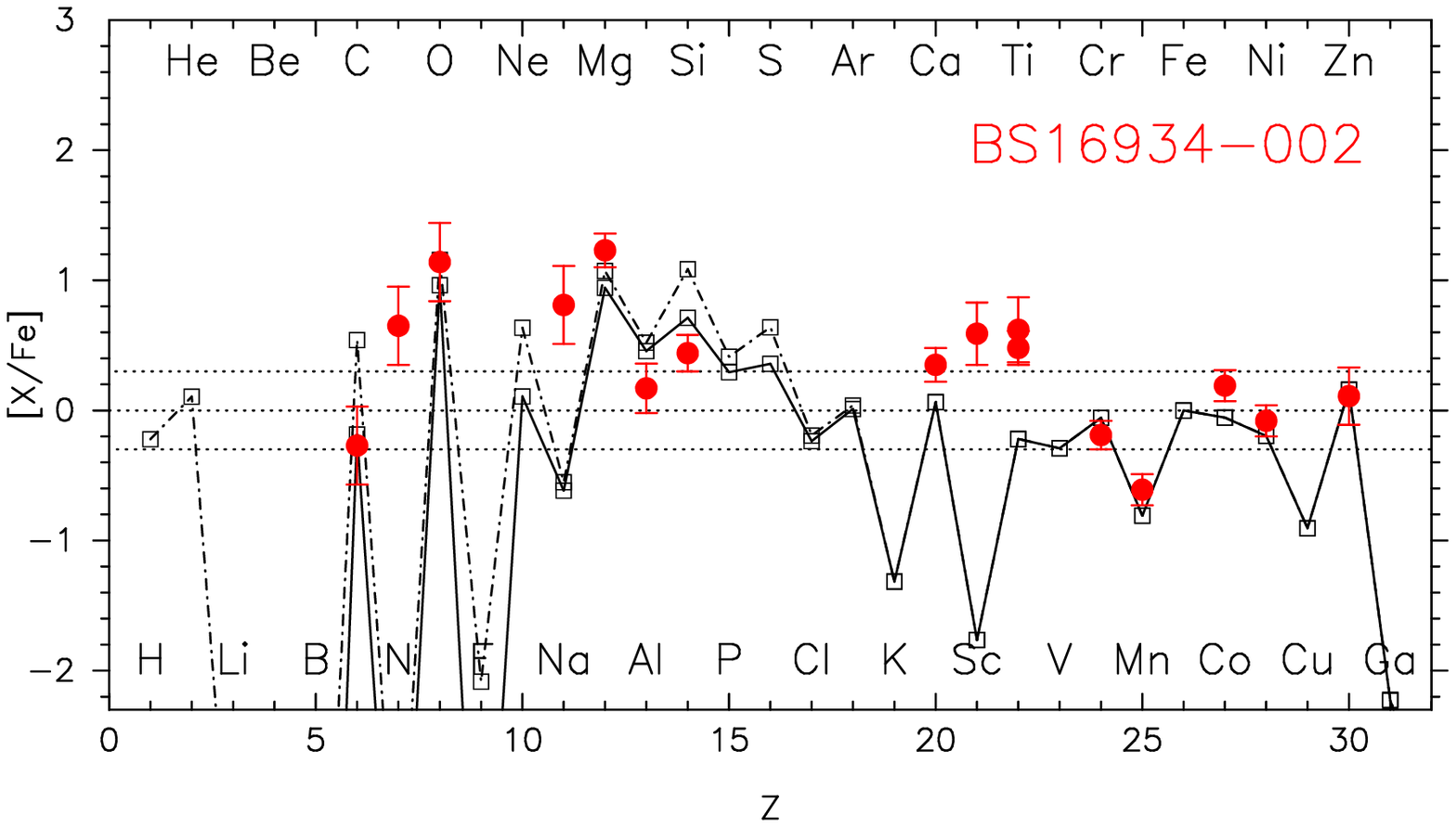} 
\end{center} 
\begin{center}
\includegraphics[width=8cm]{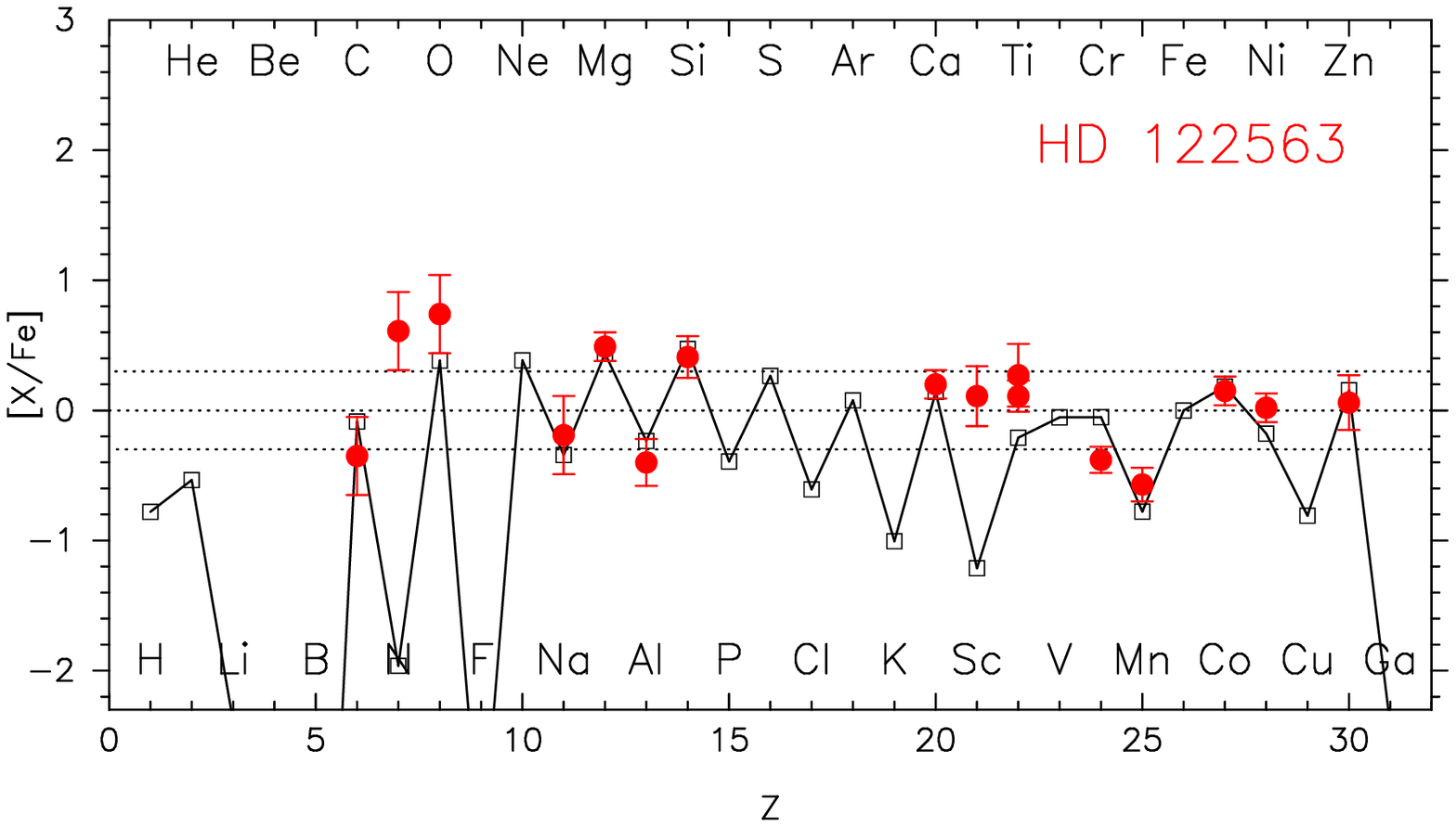} 
\end{center} 
\caption[]{Comparison of the observed abundance patterns with predicted supernova
yields. The upper panel shows the abundance pattern in {\BS} compared with the
predicted yield of 40~$M_\odot$, Population III supernova with explosion energy
of 30 $\times 10^{51}$ ergs. The abundance patterns of the explosive yields,
with and without undergoing significant mass ejection up to the CO-rich layer,
are shown by the solid and dot-dashed lines, respectively. The lower panel shows
a comparison of the abundance pattern observed in {\HD} with that of
predicted Population III supernova yields integrated over a Salpeter Initial
Mass Function with a mass range of 11-50~$M_\odot$. }
\label{fig:model}

\end{figure}

We have carried out a nucleosynthesis calculation for a supernova
whose progenitor is a 40~$M_{\odot}$, zero-metallicity (Population
III) star exploded with an explosion energy of $3\times 10^{52}$erg (a
hypernova whose explosion energy is larger than $1\times 10^{52}$ erg,
e.g., \citealt{umeda05}; Tominaga et al. 2006a).  The high
explosion energy is adopted to explain the large enhancement of Co
observed in {\BS} (Umeda \& Nomoto 2002). We assume the
mixing-fallback model, where the explosively burned materials are
mixed into the region of $M_r=$ 2.67--8.06~$M_\odot$, and only a
fraction, $f=0.025$, of the mixed materials is ejected (the ejected Fe
mass being 0.04~$M_\odot$), in order to approximately reproduce the
overall abundance pattern of {\BS} (dot-dashed line in the upper panel
of Figure~\ref{fig:model}).   Such large scale mixing and small
amount of Fe ejection in the hypernova explosion can be realized in a
jet-like explosion (Maeda \& Nomoto 2003; Tominaga et al. 2007).  In
this model, the electron fraction, $Y_e$, in the explosive ¡¡ 
Si-burning layer is modified to $Y_e=$0.5001 and 0.4997 in the
complete and incomplete Si-burning layers, respectively. This
alteration of $Y_e$ in the inner region can be realized by the
interaction of inner materials with neutrinos emitted from the
collapsed remnant (\citealt{frohlich06}; Iwamoto et al. 2006, in
preparation).  However, note that Figure ~\ref{fig:model} (dot-dashed
line) shows that our calculation predicts a large overabundance of
carbon, which is not observed for {\BS}.

The observed ratios of [C/O] $\sim -1.4$ and [C/Mg] $\sim -1.5$ in
{\BS} are extremely low among known extremely metal-poor stars.  We
find that both ratios are $-0.8$ or higher in our Population III supernova
yields in the mass range 13-50~$M_\odot$. Carbon is mainly produced by
triple-$\alpha$ reactions, and is located above the O and Mg-rich
layers. Thus, one possible modification of the above model that might
explain the lack of observed C-enhancement in {\BS} is to consider the
effects of significant mass loss from layers containing C-rich
material from its massive progenitor during its evolution. 
This is an extension of the models of SN Ib and SN Ic, where the H and
He envelopes, respectively, have been removed, possibly through
transfer across a binary. Binary-modulated stripping is suggested for
the SN IIb SN2001ig (Ryder et al. 2004, MNRAS, 349, 1093) and has
been given additional credibility up by their subsequent discovery of
a stellar-like object outliving the SN, which they believe to be the
WR companion of the SN Ib/c (Ryder et al. 2006, MNRAS, 369, 32).

\begin{figure} 
\includegraphics[width=8cm]{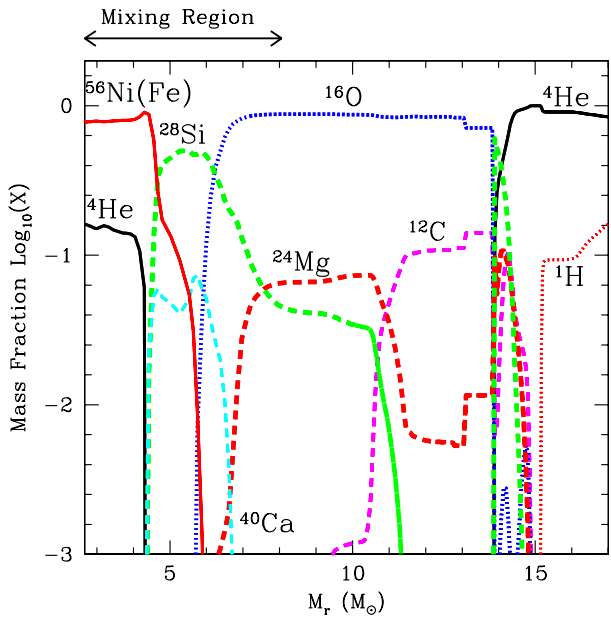} 
\caption[]{Internal abundance distribution of a 40~$M_\odot$, Population III
hypernova model.  Extensive mixing is assumed in the range
$M_r=2.67$-8.06~$M_\odot$, and the ejection fraction from the mixing
region is taken to be $f=0.025$. The ejecta have an Fe mass of 0.04~$M_\odot$. 
The region lost by an assumed significant mass ejection during the evolution of the
massive progenitor is taken to be $M_r > 11.65~M_\odot$. 
}
\label{fig:model2} 
\end{figure}

We calculated the internal abundance distribution of the
40~$M_{\odot}$, Population III hypernova model, and show the results
in Figure~\ref{fig:model2}. The loss of H, He, and CO-rich outer
layers could be explained by common-envelope evolution in a close
binary system, and later by wind-mass loss during the He and CO star
phases of the progenitor evolution.\footnote{Note that conventional
models of a Population III star have no opacity source of heavy
elements in the H envelope, hence a radiation-driven wind does not
operate, due to the low metallicity.} The solid line in the upper
panel of Figure~\ref{fig:model} shows the abundance pattern calculated
assuming significant mass ejection prior to explosion. We note here
that the ejection of He and C-rich layers may slightly affect the
internal abundance distribution even if the H-envelope is lost during
or after the core He burning phase.  However, such effect is not
included in the model calculation here.  The observed abundance ratios
of [C/Mg] and [C/Fe] in {\BS} are reproduced by this model.  Here, one
important concern is that the C ejected before the explosion might be
swept up by the shock wave, and then possibly mixed with the supernova
ejecta.  The mixing process between the circumstellar matter (CSM) and
supernova ejecta is still uncertain, and we here assume that the
mixing with the nearby CSM is inefficient.

The observed excesses of elements with odd atomic numbers (N, Na, and
Sc) are not explained by the above models. The excess of N ([N/Fe] =
+0.65) could be a result of internal processes (e.g., the first
dredge-up) during the evolution of {\BS} itself. Indeed, a similar
overabundance of N is also found in {\HD}, which has similar
atmospheric parameters, and is likely to be in a similar evolutionary
stage.  The underproduction of Sc in our supernova model is a problem
in explaining the abundances of both {\BS} and {\HD} (see below for
the latter object). This may be improved by using yields of
nucleosynthesis in the explosion under low-density conditions
\citep{umeda05}, and/or a neutrino-driven nucleosynthesis
\citep{frohlich06}. Neutrino-driven nucleosynthesis may also account
for the overproduction of Ti isotopes observed in {\BS}
\citep{pruet05,frohlich06}. The overabundance of Na in {\BS} is
possibly explained by overshooting in the convective C burning shell
during the evolution of the massive-star progenitor, as suggested by
\citet{iwamoto05}. The Na abundance ejected by a supernova is
determined by the extents of mixing and fallback regions. This is a
possible reason for the fact that the Na overabundance is found in
{\BS} but is not seen in other objects.

Another important difference between {\BS} and the two CEMP-$\alpha$ stars
CS~22949--037 and CS~29498--043 might be their iron abundances. {\BS} has a
significantly higher iron abundance ([Fe/H] $= -2.8$) than the other two stars
(both of which have [Fe/H] $\le -3.5$). The EMP stars are considered to be
formed from the ejecta of a single Population III or Population II supernova by
supernova shock compression. The metallicities of EMP stars are determined by
the ejected Fe mass, $M$(Fe), and the explosion energy $E$ ($E_{51}$: in units of
$10^{51}$~ergs) of the parent supernova as follows (a supernova-induced star
formation model, see \citealt{cioffi88,ryan96,shigeyama98, umeda05}; Tominaga et al.
2006): \begin{eqnarray} \nonumber {\rm [Fe/H]} &=& \log_{10}(M{\rm (Fe)}/M{\rm
(H)}) -\log_{10}{(X{\rm (Fe)}/X{\rm (H) })_\odot} \\ &\simeq&
\log_{10}\left({M({\rm Fe})/{M_\odot}\over{E_{51}}}\right) -C. \end{eqnarray}
where $M$(H) is the swept-up H mass, being roughly proportional to the explosion
energy $E$, and $C$ is a ``constant'' value as a function of the density and 
sound speed of the CSM. 

In the context of a supernova-induced star formation model, CS~29498--043
([Fe/H]$=-3.75$, \citealt{aoki04}) can be reproduced with ``constant'' $C=0.8$
(a model with $M{\rm (Fe) }=0.0011M_\odot$ and $E_{51}=1$, as shown in Figure~9a
of \citealt{umeda05}). If the same constant $C$ is applied to {\BS}, the
predicted [Fe/H] of {\BS} would be [Fe/H]$=-3.7$, similar to CS~29498--043,
because of the large explosion energy ($E_{51}=30$), in spite of the large
amount of Fe ejection ($M{\rm (Fe) }=0.04M_\odot$). Therefore, {\BS} might have
been formed from the ejecta of a Population II supernova that already had a
higher metallicity, e.g., [Fe/H]$\sim-3$. However, the star-formation process is
affected by the uncertain distribution of the surrounding gas in the early
Galaxy. Thus, there remains the possibility that Population III hypernovae form a
portion of the [Fe/H] $\sim-3$ stars.

On the other hand, {\HD} has a metallicity [Fe/H] $= -2.6$, thus is a likely
second- or later-generation star \citep[e.g.,][]{tumlinson06}. Therefore, we
model the abundance pattern of {\HD} using the yields integrated over a mass
range of 11-50~$M_\odot$ with a Salpeter Initial Mass Function, as shown in the
lower panel of Figure~\ref{fig:model}. The Population III supernova models
employed are ($M_{\rm MS}$, $E_{51}$) = (13,1), (15,1), (18,1), (20,10),
(25,10), (30,20), (40,30), and (50,40) (\citealt{kobayashi06}; Nomoto et al.
2006; Tominaga et al. 2006), where $M_{\rm MS}$ is the main-sequence mass of
the progenitor in solar masses. These models, with $M_{\rm MS} > 20M_\odot$, are
assumed to explode as hypernovae and undergo mixing-fallback. Small variations
of $Y_e$ are assumed, as in the model for {\BS} discussed above. We note that
the yields of Population III supernovae can be used even if {\HD} is actually formed
from the ejecta of Population II supernovae, because the explosive yield of a
Population III model is not so different from that of a model with [Fe/H] $=-3$ (Tominaga et
al. 2006).

We have considered supernova explosions in an interacting binary system to
explain the significant mass loss from the massive progenitor of {\BS}. Our
interpretation, however, raises important questions as to why {\BS} has such a
peculiar composition, and why it alone {\BS} might require binarity in order to
explain the observed abundance pattern. More investigations
are clearly necessary to resolve these questions.

\subsection{The Spread of Mg/Fe in Extremely Metal-Poor Stars}\label{sec:mgfe}

The small spread of the observed abundance ratios of the $\alpha$ elements (e.g., [Mg/Fe]) in
very metal-poor stars has been emphasized by recent work
\citep[e.g. ][]{Cayreletal:2004, arnone05}. The bulk of halo stars
exhibit moderate overabundances of Mg, with star-to-star scatter as small as
0.1~dex, while much larger scatter is found for other elements, such as the
neutron-capture elements. One possible interpretation for the small spread of
the [Mg/Fe] ratios is that the Galactic halo was well mixed even in the earliest
phases of its evolution \citep{francois04}. The Mg abundance of {\BS} clearly
deviates from the trend found in halo stars, suggesting that this star has a
rather different nucleosynthesis history than the majority of halo objects.
Other stars that exhibit departures from the abundance trend of the $\alpha$ elements
are the ADMP stars, with large underabundances of Mg \citep[e.g. ][]{ivans03};
these abundance patterns are not simply explained by any combination of the yields of Type
Ia and II supernovae. Understanding the abundance patterns of AEMP stars, as well
as those of ADMP stars, remains a challenge for studies of the
nucleosynthesis processes that pertain to low-mass stars in the very early Galaxy.

Although recent studies based on high-resolution spectroscopy have revealed the
chemical compositions of a significant number of EMP stars, much larger studies
are required in order to better understand the observed abundance trends, and
the scatter about these trends, in halo stars. In particular, for the lowest
metallicity range ($-4.0\lesssim$ [Fe/H] $\lesssim-3.5$), at least two
Mg-enhanced objects are known among the relatively small number of stars (10--20
objects) investigated do date, suggesting that a large scatter of the Mg
abundance ratios may exist in this metallicity regime. The AEMP star {\BS} might
provide the first clues for developing an understanding of the relationship
between the abundance scatter of $\alpha$ elements in the lowest metallicity
range and the clear trend found at relatively higher metallicity. 

\acknowledgments

W.A. and K.N. are supported by a Grant-in-Aid for Science Research
from JSPS (grant 152040109, 18104003). T.C.B. acknowledges partial
funding for this work from grant AST 04-06784, as well as from grant
PHY 02-16783: Physics Frontiers Center/Joint Institute for Nuclear
Astrophysics (JINA), both awarded by the U.S.  National Science
Foundation. K.N. has been supported in part by the Grant-in-Aid for
Scientific Research (17030005, 17033002, 18540231) and the 21st
Century COE Program (QUEST) from the JSPS and MEXT of Japan. N.I. has
been supported by the Grant-in-Aid for Young Scientists (B)
(17740163). N.T. is supported through the JSPS Research Fellowship for
Young Scientists. J.E.N. acknowledges support from the Australian
Research Council under grant DP0342613.


\clearpage
\begin{deluxetable}{@{}l@{\extracolsep{\fill}}c@{\extracolsep{\fill}}c@{\extracolsep{\fill}}c@{\extracolsep{\fill}}c@{\extracolsep{\fill}}c@{\extracolsep{\fill}}c}
\tablewidth{0pt}
\tablecaption{\label{tab:obs} PROGRAM STARS AND OBSERVATIONS}
\tablehead{
Star    & Wavelength &  Exp.\tablenotemark{a} & S/N\tablenotemark{b} &  Obs. date (JD) & Radial velocity ~ & \\
        & (~{\AA})   &     (minutes)                   &                      &                 & (km s$^{-1}$)     &    
}
\startdata
{\BS} & 4030--6800 & 60 (2) & 100/1 & 31 May, 2004  (2453158)  & $82.5 \pm 0.3$   &   \\
{\HD} & 4030--6800 & 2 (2) & 300/1  & 19 June, 2005  (2453541)  & $-26.0 \pm 0.2$   &   
\enddata
\tablenotetext{a}{Exposure time (number of exposures).}
\tablenotetext{b}{$S/N$ ratio per pixel (0.18{\kms}) estimated from th photon counts at 4500~{\AA}}
\end{deluxetable}

\begin{deluxetable}{lcccccc}
\tablewidth{0pt}
\tablecaption{EQUIVALENT WIDTHS\label{tab:ew}} 
\tablehead{
Species    & wavelength & {\loggf} & L.E.P. & {\HD} & {\BS}  & Remarks 
}
\startdata
     O I  &  6300.300 &   -9.820 &  0.000  &    7.2 &   14.3 & \\
     O I  &  6363.776 &  -10.303 &  0.020  &    2.2 &    5.0 & \\
     Na I  &  5682.633 &  -0.700 &   2.102 &    1.3 &    8.7 & \\
     Na I  &  5688.204 &  -0.412 &   2.104 &    2.8 &   17.7 & \\
     Na I  &  5889.951 &   0.101 &   0.000 &    ... &  225.6 & \\
     Na I  &  5895.924 &  -0.197 &   0.000 &    ... &  212.0 & \\
     Mg I  &  3829.355 &  -0.210 &   2.709 &  182.1 &    ... & \\
     Mg I  &  3986.753 &  -1.030 &   4.346 &   35.4 &    ... & \\
     Mg I  &  4057.505 &  -0.890 &   4.346 &    ... &   72.8 & \\
     Mg I  &  4167.271 &  -0.770 &   4.346 &   54.4 &   84.6 & \\
     Mg I  &  4571.096 &  -5.688 &   0.000 &   84.8 &  113.3 & \\
     Mg I  &  4702.991 &  -0.520 &   4.346 &   73.8 &  106.8 & \\
     Mg I  &  5172.685 &  -0.380 &   2.712 &  201.7 &    ... & \\
     Mg I  &  5183.604 &  -0.160 &   2.717 &  221.8 &    ... & \\
     Mg I  &  5528.404 &  -0.490 &   4.346 &   77.6 &  112.9 & \\
     Mg I  &  5711.088 &  -1.720 &   4.346 &   11.7 &   31.7 & \\
     Al I  &  3944.006 &  -0.644 &   0.000 &  163.0 &    ... & \\
     Al I  &  3961.520 &  -0.340 &   0.014 &  139.0 &  153.3 & \\
     Si I  &  3905.523 &  -0.980 &   1.909 &  196.6 &    ... & \\
     Si I  &  4102.936 &  -2.910 &   1.909 &   88.8 &   78.5 & \\
     Ca I  &  4226.728 &   0.244 &   0.000 &    ... &  201.5 & \\
     Ca I  &  4283.010 &  -0.220 &   1.886 &   62.7 &   68.3 & \\
     Ca I  &  4318.652 &  -0.210 &   1.899 &   53.7 &   52.7 & \\
     Ca I  &  4425.441 &  -0.360 &   1.879 &   49.1 &   45.9 & \\
     Ca I  &  4435.688 &  -0.520 &   1.886 &   41.5 &   38.2 & \\
     Ca I  &  4454.781 &   0.260 &   1.899 &    ... &   83.3 & \\
     Ca I  &  4455.887 &  -0.530 &   1.899 &   39.8 &   43.9 & \\
     Ca I  &  5262.244 &  -0.471 &   2.521 &   34.2 &   44.6 & \\
     Ca I  &  5581.971 &  -0.555 &   2.523 &   13.1 &   12.0 & \\
     Ca I  &  5588.757 &   0.358 &   2.526 &   49.6 &   49.4 & \\
     Ca I  &  5590.120 &  -0.571 &   2.521 &   12.3 &   12.0 & \\
     Ca I  &  5594.468 &   0.097 &   2.523 &   36.7 &   36.2 & \\
     Ca I  &  5598.487 &  -0.087 &   2.521 &   29.9 &   28.6 & \\
     Ca I  &  5601.285 &  -0.523 &   2.526 &   13.1 &   10.7 & \\
     Ca I  &  5857.452 &   0.240 &   2.933 &   24.3 &   25.5 & \\
     Ca I  &  6102.722 &  -0.770 &   1.879 &   36.8 &   39.4 & \\
     Ca I  &  6122.219 &  -0.320 &   1.886 &   68.5 &   64.4 & \\
     Ca I  &  6169.055 &  -0.797 &   2.523 &    9.4 &   10.3 & \\
     Ca I  &  6169.559 &  -0.478 &   2.526 &   16.3 &   14.5 & \\
     Ca I  &  6439.073 &   0.390 &   2.526 &   58.8 &   61.4 & \\
     Ca I  &  6449.810 &  -0.502 &   2.521 &   16.5 &   16.0 & \\
     Ca I  &  6499.649 &  -0.818 &   2.523 &    9.4 &   10.7 & \\
     Ca I  &  6717.685 &  -0.524 &   2.709 &   11.1 &   13.0 & \\
     Sc I  &  4995.002 &  -0.880 &   2.111 &    1.0 &    ... & \\
     Ti I  &  3904.784 &   0.030 &   0.900 &   29.6 &   35.3 & \\
     Ti I  &  3924.526 &  -0.881 &   0.021 &   35.1 &   48.0 & \\
     Ti I  &  3989.759 &  -0.142 &   0.021 &   71.7 &   86.5 & \\
     Ti I  &  3998.637 &   0.000 &   0.048 &   71.5 &   84.2 & \\
     Ti I  &  4008.928 &  -1.016 &   0.021 &   29.5 &   41.4 & \\
     Ti I  &  4512.733 &  -0.424 &   0.836 &   14.1 &   27.4 & \\
     Ti I  &  4533.239 &   0.532 &   0.848 &   52.9 &   67.4 & \\
     Ti I  &  4534.775 &   0.336 &   0.836 &   42.2 &   57.2 & \\
     Ti I  &  4535.567 &   0.120 &   0.826 &   36.2 &   50.1 & \\
     Ti I  &  4544.687 &  -0.520 &   0.818 &   14.9 &   22.3 & \\
     Ti I  &  4548.763 &  -0.298 &   0.826 &   17.9 &   28.0 & \\
     Ti I  &  4681.908 &  -1.015 &   0.048 &   29.5 &   48.2 & \\
     Ti I  &  4981.730 &   0.560 &   0.848 &   62.2 &    ... & \\
     Ti I  &  4991.066 &   0.436 &   0.836 &   58.1 &   73.2 & \\
     Ti I  &  4999.501 &   0.306 &   0.826 &   50.2 &   66.6 & \\
     Ti I  &  5007.206 &   0.168 &   0.818 &   52.2 &   64.0 & \\
     Ti I  &  5009.645 &  -2.203 &   0.021 &    ... &    5.7 & \\
     Ti I  &  5016.160 &  -0.516 &   0.848 &   12.7 &   23.7 & \\
     Ti I  &  5020.024 &  -0.358 &   0.836 &   18.3 &   30.1 & \\
     Ti I  &  5024.843 &  -0.546 &   0.818 &   13.6 &   21.5 & \\
     Ti I  &  5035.902 &   0.260 &   1.460 &   16.4 &   26.3 & \\
     Ti I  &  5036.463 &   0.186 &   1.443 &   11.2 &   19.9 & \\
     Ti I  &  5038.396 &   0.069 &   1.430 &    ... &   18.7 & \\
     Ti I  &  5039.960 &  -1.130 &   0.020 &   29.9 &   51.0 & \\
     Ti I  &  5064.651 &  -0.935 &   0.048 &   35.4 &   53.1 & \\
     Ti I  &  5173.740 &  -1.062 &   0.000 &   33.4 &   50.3 & \\
     Ti I  &  5192.969 &  -0.948 &   0.021 &   38.3 &   55.1 & \\
     Ti I  &  5210.384 &  -0.828 &   0.048 &   42.6 &   62.6 & \\
     Ti I  &  6258.099 &  -0.299 &   1.443 &    5.4 &    8.6 & \\
      V I  &  4379.230 &   0.550 &   0.301 &   30.6 &   48.6 & \\
      V I  &  4786.499 &   0.110 &   2.074 &   16.0 &    9.8 & \\
      V I  &  5195.395 &  -0.120 &   2.278 &   11.8 &    7.2 & \\
      V I  &  6008.648 &  -2.340 &   1.183 &    6.1 &    4.0 & \\
     Cr I  &  3908.762 &  -1.000 &   1.004 &   21.6 &    ... & \\
     Cr I  &  4254.332 &  -0.114 &   0.000 &  113.1 &  118.6 & \\
     Cr I  &  4274.796 &  -0.231 &   0.000 &  112.9 &  109.3 & \\
     Cr I  &  4337.566 &  -1.112 &   0.968 &   19.3 &   26.4 & \\
     Cr I  &  4558.650 &  -0.660 &   4.070 &   19.0 &   21.8 & \\
     Cr I  &  4580.056 &  -1.650 &   0.941 &   16.7 &   16.2 & \\
     Cr I  &  4588.200 &  -0.630 &   4.070 &   12.7 &   15.0 & \\
     Cr I  &  4600.752 &  -1.260 &   1.004 &   16.6 &   19.4 & \\
     Cr I  &  4616.137 &  -1.190 &   0.983 &   19.9 &   22.1 & \\
     Cr I  &  4626.188 &  -1.320 &   0.968 &   16.8 &   15.6 & \\
     Cr I  &  4646.174 &  -0.700 &   1.030 &   35.6 &    ... & \\
     Cr I  &  4651.285 &  -1.460 &   0.983 &   11.8 &   13.4 & \\
     Cr I  &  4652.158 &  -1.030 &   1.004 &   24.6 &   29.3 & \\
     Cr I  &  5206.039 &   0.019 &   0.941 &   85.8 &   91.3 & \\
     Cr I  &  5247.564 &  -1.640 &   0.961 &    9.8 &   13.6 & \\
     Mn I  &  3823.508 &   0.058 &   2.143 &   16.6 &    ... & \\
     Mn I  &  4030.753 &  -0.470 &   0.000 &  138.4 &  122.9 & \\
     Mn I  &  4033.062 &  -0.618 &   0.000 &  125.6 &  116.6 & \\
     Mn I  &  4034.483 &  -0.811 &   0.000 &  117.4 &  112.3 & \\
     Mn I  &  4041.357 &   0.285 &   2.114 &   45.3 &   27.4 & \\
     Mn I  &  4754.048 &  -0.086 &   2.282 &   19.2 &   13.9 & \\
     Mn I  &  4783.432 &   0.042 &   2.298 &   23.6 &   16.1 & \\
     Mn I  &  4823.528 &   0.144 &   2.319 &   25.9 &   16.1 & \\
     Fe I  &  3787.880 &  -0.859 &   1.011 &  144.9 &  134.0 & \\
     Fe I  &  3805.342 &   0.310 &   3.301 &   66.2 &   54.8 & \\
     Fe I  &  3815.840 &   0.226 &   1.485 &  174.2 &  161.7 & \\
     Fe I  &  3827.823 &   0.062 &   1.557 &  159.9 &  152.4 & \\
     Fe I  &  3839.256 &  -0.330 &   3.047 &   50.4 &   55.0 & \\
     Fe I  &  3840.437 &  -0.506 &   0.990 &  165.4 &  146.4 & \\
     Fe I  &  3841.048 &  -0.065 &   1.608 &  144.1 &  134.3 & \\
     Fe I  &  3845.169 &  -1.390 &   2.424 &   44.7 &   35.6 & \\
     Fe I  &  3846.800 &  -0.020 &   3.251 &   55.0 &   51.0 & \\
     Fe I  &  3849.980 &  -0.863 &   1.010 &  149.5 &  131.7 & \\
     Fe I  &  3850.818 &  -1.745 &   0.990 &    ... &  117.6 & \\
     Fe I  &  3852.573 &  -1.180 &   2.176 &   71.5 &   63.6 & \\
     Fe I  &  3856.372 &  -1.286 &   0.052 &  182.2 &  178.9 & \\
     Fe I  &  3863.741 &  -1.430 &   2.692 &   39.8 &   31.4 & \\
     Fe I  &  3865.523 &  -0.982 &   1.011 &  149.0 &  141.0 & \\
     Fe I  &  3867.216 &  -0.450 &   3.017 &   46.7 &   41.6 & \\
     Fe I  &  3885.510 &  -1.090 &   2.424 &   58.6 &   42.5 & \\
     Fe I  &  3899.707 &  -1.531 &   0.087 &  172.1 &  156.8 & \\
     Fe I  &  3902.946 &  -0.466 &   1.557 &  136.8 &  118.9 & \\
     Fe I  &  3917.181 &  -2.155 &   0.990 &  108.4 &    ... & \\
     Fe I  &  3920.258 &  -1.746 &   0.121 &  161.9 &  156.0 & \\
     Fe I  &  3922.912 &  -1.651 &   0.052 &  172.7 &  163.6 & \\
     Fe I  &  3940.878 &  -2.600 &   0.958 &   83.5 &   84.1 & \\
     Fe I  &  3949.953 &  -1.250 &   2.176 &   67.1 &   73.5 & \\
     Fe I  &  3977.741 &  -1.120 &   2.198 &   74.8 &   71.5 & \\
     Fe I  &  4001.661 &  -1.900 &   2.176 &   37.7 &    ... & \\
     Fe I  &  4005.242 &  -0.610 &   1.557 &  132.2 &  119.5 & \\
     Fe I  &  4007.272 &  -1.280 &   2.759 &   28.6 &   31.3 & \\
     Fe I  &  4014.531 &  -0.590 &   3.047 &   71.8 &    ... & \\
     Fe I  &  4021.866 &  -0.730 &   2.759 &   54.2 &    ... & \\
     Fe I  &  4032.627 &  -2.380 &   1.485 &   54.0 &   45.3 & \\
     Fe I  &  4044.609 &  -1.220 &   2.832 &   32.3 &    ... & \\
     Fe I  &  4058.217 &  -1.110 &   3.211 &   30.7 &   26.4 & \\
     Fe I  &  4062.441 &  -0.860 &   2.845 &   45.7 &   43.9 & \\
     Fe I  &  4063.594 &   0.060 &   1.557 &  166.0 &  151.8 & \\
     Fe I  &  4067.271 &  -1.419 &   2.559 &   41.4 &   32.5 & \\
     Fe I  &  4067.978 &  -0.470 &   3.211 &   43.4 &   35.3 & \\
     Fe I  &  4070.769 &  -0.790 &   3.241 &   27.1 &   22.4 & \\
     Fe I  &  4071.738 &  -0.022 &   1.608 &  156.2 &  144.1 & \\
     Fe I  &  4073.763 &  -0.900 &   3.266 &   22.2 &    ... & \\
     Fe I  &  4079.838 &  -1.360 &   2.858 &   22.7 &   18.7 & \\
     Fe I  &  4095.970 &  -1.480 &   2.588 &    ... &   27.4 & \\
     Fe I  &  4098.176 &  -0.880 &   3.241 &   25.5 &    ... & \\
     Fe I  &  4109.057 &  -1.560 &   3.292 &   11.7 &   15.3 & \\
     Fe I  &  4109.802 &  -0.940 &   2.845 &   45.0 &   44.9 & \\
     Fe I  &  4114.445 &  -1.300 &   2.832 &   30.1 &   26.5 & \\
     Fe I  &  4120.207 &  -1.270 &   2.990 &   24.3 &   16.7 & \\
     Fe I  &  4121.802 &  -1.450 &   2.832 &   24.4 &   20.9 & \\
     Fe I  &  4132.899 &  -1.010 &   2.845 &   43.3 &   38.9 & \\
     Fe I  &  4136.998 &  -0.450 &   3.415 &   26.7 &   25.1 & \\
     Fe I  &  4139.927 &  -3.629 &   0.990 &    ... &   28.9 & \\
     Fe I  &  4143.414 &  -0.200 &   3.047 &   67.1 &   58.4 & \\
     Fe I  &  4143.868 &  -0.510 &   1.557 &  135.2 &  123.9 & \\
     Fe I  &  4147.669 &  -2.104 &   1.485 &   80.4 &   74.2 & \\
     Fe I  &  4152.169 &  -3.232 &   0.958 &   60.8 &   49.9 & \\
     Fe I  &  4153.899 &  -0.320 &   3.397 &   40.7 &   32.8 & \\
     Fe I  &  4154.499 &  -0.690 &   2.832 &   54.0 &   43.7 & \\
     Fe I  &  4154.805 &  -0.400 &   3.368 &   39.8 &   32.9 & \\
     Fe I  &  4156.799 &  -0.810 &   2.832 &   59.3 &   51.6 & \\
     Fe I  &  4157.779 &  -0.400 &   3.417 &   37.3 &   38.0 & \\
     Fe I  &  4158.793 &  -0.670 &   3.430 &   22.4 &   17.4 & \\
     Fe I  &  4175.636 &  -0.830 &   2.845 &   52.1 &   47.5 & \\
     Fe I  &  4181.755 &  -0.370 &   2.832 &   75.4 &   68.8 & \\
     Fe I  &  4182.382 &  -1.180 &   3.017 &   24.5 &   16.6 & \\
     Fe I  &  4184.892 &  -0.870 &   2.832 &   45.2 &   39.4 & \\
     Fe I  &  4187.039 &  -0.548 &   2.450 &   89.5 &   80.7 & \\
     Fe I  &  4191.430 &  -0.670 &   2.469 &   84.9 &    ... & \\
     Fe I  &  4195.329 &  -0.490 &   3.332 &   44.2 &   39.0 & \\
     Fe I  &  4196.208 &  -0.700 &   3.397 &   23.1 &   19.8 & \\
     Fe I  &  4199.095 &   0.160 &   3.047 &   80.0 &   75.6 & \\
     Fe I  &  4216.184 &  -3.356 &   0.000 &  115.3 &  108.5 & \\
     Fe I  &  4222.213 &  -0.967 &   2.450 &    ... &   69.8 & \\
     Fe I  &  4227.426 &   0.270 &   3.332 &   91.9 &    ... & \\
     Fe I  &  4233.603 &  -0.604 &   2.482 &   85.3 &   80.6 & \\
     Fe I  &  4238.810 &  -0.230 &   3.397 &   44.2 &   39.8 & \\
     Fe I  &  4247.426 &  -0.240 &   3.368 &   50.6 &   44.6 & \\
     Fe I  &  4250.119 &  -0.405 &   2.469 &   94.4 &   81.0 & \\
     Fe I  &  4250.787 &  -0.710 &   1.557 &  129.1 &    ... & \\
     Fe I  &  4260.474 &   0.080 &   2.399 &  119.4 &  111.7 & \\
     Fe I  &  4271.153 &  -0.349 &   2.450 &  103.8 &   96.4 & \\
     Fe I  &  4271.761 &  -0.164 &   1.485 &  163.7 &  152.4 & \\
     Fe I  &  4282.403 &  -0.780 &   2.176 &   92.1 &   84.4 & \\
     Fe I  &  4325.762 &   0.010 &   1.608 &  162.0 &  150.4 & \\
     Fe I  &  4337.046 &  -1.695 &   1.557 &   94.5 &    ... & \\
     Fe I  &  4352.735 &  -1.290 &   2.223 &   77.7 &   72.1 & \\
     Fe I  &  4375.930 &  -3.031 &   0.000 &  124.8 &  122.2 & \\
     Fe I  &  4407.709 &  -1.970 &   2.176 &   59.0 &    ... & \\
     Fe I  &  4422.568 &  -1.110 &   2.845 &   45.6 &    ... & \\
     Fe I  &  4430.614 &  -1.659 &   2.223 &   54.6 &   46.2 & \\
     Fe I  &  4442.339 &  -1.255 &   2.198 &   78.7 &   75.3 & \\
     Fe I  &  4443.194 &  -1.040 &   2.858 &   38.1 &   31.7 & \\
     Fe I  &  4447.717 &  -1.342 &   2.223 &   73.8 &   65.1 & \\
     Fe I  &  4454.381 &  -1.300 &   2.832 &    ... &   29.6 & \\
     Fe I  &  4459.118 &  -1.279 &   2.176 &   92.3 &   76.4 & \\
     Fe I  &  4461.653 &  -3.210 &   0.087 &  119.0 &  112.4 & \\
     Fe I  &  4466.552 &  -0.600 &   2.832 &   86.3 &   77.0 & \\
     Fe I  &  4476.019 &  -0.820 &   2.845 &   67.6 &   57.3 & \\
     Fe I  &  4484.220 &  -0.860 &   3.603 &   17.2 &   16.6 & \\
     Fe I  &  4489.739 &  -3.966 &   0.121 &   87.4 &   80.6 & \\
     Fe I  &  4490.083 &  -1.580 &   3.017 &    9.6 &    6.6 & \\
     Fe I  &  4494.563 &  -1.136 &   2.198 &   84.6 &   77.9 & \\
     Fe I  &  4528.614 &  -0.822 &   2.176 &  102.8 &   96.4 & \\
     Fe I  &  4531.148 &  -2.155 &   1.485 &   84.9 &   74.8 & \\
     Fe I  &  4592.651 &  -2.449 &   1.557 &   68.3 &   63.0 & \\
     Fe I  &  4602.941 &  -2.210 &   1.485 &   82.5 &   78.1 & \\
     Fe I  &  4611.185 &  -2.720 &   2.851 &   15.3 &    ... & \\
     Fe I  &  4632.912 &  -2.913 &   1.608 &   34.4 &   28.0 & \\
     Fe I  &  4647.435 &  -1.350 &   2.949 &   24.3 &   21.1 & \\
     Fe I  &  4678.846 &  -0.830 &   3.603 &   17.9 &   16.4 & \\
     Fe I  &  4691.411 &  -1.520 &   2.990 &   20.1 &   20.5 & \\
     Fe I  &  4707.274 &  -1.080 &   3.241 &   28.5 &    ... & \\
     Fe I  &  4710.283 &  -1.610 &   3.018 &   15.6 &   14.8 & \\
     Fe I  &  4733.591 &  -2.990 &   1.485 &   40.3 &   35.3 & \\
     Fe I  &  4736.772 &  -0.750 &   3.211 &   44.1 &   36.0 & \\
     Fe I  &  4786.807 &  -1.610 &   3.017 &   15.6 &    ... & \\
     Fe I  &  4789.650 &  -0.960 &   3.547 &   12.1 &    8.7 & \\
     Fe I  &  4859.741 &  -0.760 &   2.876 &    ... &   53.2 & \\
     Fe I  &  4871.318 &  -0.360 &   2.865 &   79.4 &   73.2 & \\
     Fe I  &  4872.138 &  -0.570 &   2.882 &   68.1 &   59.7 & \\
     Fe I  &  4890.755 &  -0.390 &   2.876 &   80.2 &   70.6 & \\
     Fe I  &  4891.492 &  -0.110 &   2.851 &   93.5 &   84.7 & \\
     Fe I  &  4903.310 &  -0.930 &   2.882 &   50.7 &   45.3 & \\
     Fe I  &  4918.994 &  -0.340 &   2.865 &   81.4 &   75.8 & \\
     Fe I  &  4920.502 &   0.070 &   2.833 &  104.1 &   96.6 & \\
     Fe I  &  4924.770 &  -2.256 &   2.279 &   30.3 &   22.6 & \\
     Fe I  &  4938.814 &  -1.080 &   2.876 &   42.3 &   35.0 & \\
     Fe I  &  4939.687 &  -3.340 &   0.859 &   71.1 &   61.9 & \\
     Fe I  &  4946.388 &  -1.170 &   3.368 &   15.9 &    ... & \\
     Fe I  &  4966.088 &  -0.870 &   3.332 &   31.3 &   26.6 & \\
     Fe I  &  4973.102 &  -0.950 &   3.960 &    8.8 &    ... & \\
     Fe I  &  4994.130 &  -2.956 &   0.915 &   82.2 &   74.5 & \\
     Fe I  &  5001.870 &   0.050 &   3.880 &   31.6 &   25.0 & \\
     Fe I  &  5006.119 &  -0.610 &   2.833 &   71.6 &   66.2 & \\
     Fe I  &  5012.068 &  -2.642 &   0.859 &  110.5 &  105.3 & \\
     Fe I  &  5014.942 &  -0.300 &   3.943 &   23.3 &   18.0 & \\
     Fe I  &  5022.236 &  -0.530 &   3.984 &   13.8 &   12.6 & \\
     Fe I  &  5027.225 &  -1.890 &   3.640 &    9.9 &    ... & \\
     Fe I  &  5028.127 &  -1.120 &   3.573 &    9.5 &    4.2 & \\
     Fe I  &  5041.072 &  -3.090 &   0.958 &   86.1 &   81.0 & \\
     Fe I  &  5041.756 &  -2.200 &   1.485 &   94.3 &   87.3 & \\
     Fe I  &  5044.210 &  -2.040 &   2.850 &    9.6 &    ... & \\
     Fe I  &  5049.820 &  -1.344 &   2.279 &   74.4 &   66.5 & \\
     Fe I  &  5051.635 &  -2.795 &   0.915 &  100.5 &   92.5 & \\
     Fe I  &  5060.080 &  -5.460 &   0.000 &   19.4 &   16.2 & \\
     Fe I  &  5068.766 &  -1.040 &   2.940 &   39.7 &   31.8 & \\
     Fe I  &  5074.749 &  -0.200 &   4.220 &   17.5 &   14.7 & \\
     Fe I  &  5079.224 &  -2.067 &   2.198 &   43.3 &   36.9 & \\
     Fe I  &  5079.740 &  -3.220 &   0.990 &   70.0 &    ... & \\
     Fe I  &  5080.950 &  -3.090 &   3.267 &   17.1 &    ... & \\
     Fe I  &  5083.339 &  -2.958 &   0.958 &   86.6 &   79.3 & \\
     Fe I  &  5090.770 &  -0.360 &   4.260 &    7.1 &    4.8 & \\
     Fe I  &  5098.697 &  -2.030 &   2.176 &   54.7 &   47.8 & \\
     Fe I  &  5123.720 &  -3.068 &   1.011 &   76.8 &    ... & \\
     Fe I  &  5125.117 &  -0.140 &   4.220 &   17.5 &   14.0 & \\
     Fe I  &  5127.359 &  -3.307 &   0.915 &   70.4 &   65.8 & \\
     Fe I  &  5131.468 &  -2.510 &   2.223 &   20.6 &   16.2 & \\
     Fe I  &  5133.689 &   0.140 &   4.178 &   31.0 &   26.8 & \\
     Fe I  &  5137.382 &  -0.400 &   4.178 &   14.4 &   11.5 & \\
     Fe I  &  5141.739 &  -1.964 &   2.424 &   19.3 &   18.1 & \\
     Fe I  &  5142.929 &  -3.080 &   0.958 &   82.5 &   73.1 & \\
     Fe I  &  5150.839 &  -3.003 &   0.990 &   73.4 &   61.6 & \\
     Fe I  &  5151.911 &  -3.322 &   1.011 &   61.7 &   54.0 & \\
     Fe I  &  5162.273 &   0.020 &   4.178 &   26.2 &   19.5 & \\
     Fe I  &  5166.282 &  -4.195 &   0.000 &   92.2 &   85.4 & \\
     Fe I  &  5171.597 &  -1.793 &   1.485 &  108.3 &  104.3 & \\
     Fe I  &  5191.455 &  -0.550 &   3.038 &   60.4 &   51.0 & \\
     Fe I  &  5192.344 &  -0.420 &   2.998 &   69.8 &   61.4 & \\
     Fe I  &  5194.942 &  -2.090 &   1.557 &   89.9 &   85.2 & \\
     Fe I  &  5198.711 &  -2.135 &   2.223 &   37.1 &   31.4 & \\
     Fe I  &  5202.336 &  -1.838 &   2.176 &   59.2 &   48.8 & \\
     Fe I  &  5204.583 &  -4.332 &   0.087 &  128.3 &    ... & \\
     Fe I  &  5216.274 &  -2.150 &   1.608 &   82.8 &   78.5 & \\
     Fe I  &  5217.390 &  -1.070 &   3.211 &   22.7 &   19.7 & \\
     Fe I  &  5225.525 &  -4.789 &   0.110 &   44.8 &   39.1 & \\
     Fe I  &  5232.940 &  -0.060 &   2.940 &   92.2 &   85.3 & \\
     Fe I  &  5242.491 &  -0.970 &   3.634 &   11.7 &    9.9 & \\
     Fe I  &  5247.050 &  -4.946 &   0.087 &   36.7 &   33.2 & \\
     Fe I  &  5250.210 &  -4.938 &   0.121 &   35.4 &   28.5 & \\
     Fe I  &  5250.646 &  -2.180 &   2.198 &   40.9 &   33.2 & \\
     Fe I  &  5254.956 &  -4.764 &   0.110 &    ... &   42.2 & \\
     Fe I  &  5263.305 &  -0.879 &   3.266 &   27.8 &   19.9 & \\
     Fe I  &  5269.537 &  -1.321 &   0.859 &  165.8 &  156.2 & \\
     Fe I  &  5281.790 &  -0.830 &   3.038 &   44.8 &   42.3 & \\
     Fe I  &  5283.621 &  -0.432 &   3.241 &   52.1 &   45.9 & \\
     Fe I  &  5302.300 &  -0.720 &   3.283 &    ... &   28.2 & \\
     Fe I  &  5307.361 &  -2.987 &   1.608 &   33.7 &   28.8 & \\
     Fe I  &  5324.179 &  -0.103 &   3.211 &   72.4 &   65.0 & \\
     Fe I  &  5328.039 &  -1.466 &   0.915 &  159.7 &  149.9 & \\
     Fe I  &  5328.531 &  -1.850 &   1.557 &  107.1 &   96.6 & \\
     Fe I  &  5332.900 &  -2.780 &   1.557 &   44.6 &   38.5 & \\
     Fe I  &  5501.465 &  -3.050 &   0.958 &   87.7 &   79.9 & \\
     Fe I  &  5506.778 &  -2.797 &   0.990 &   99.8 &   91.4 & \\
     Fe I  &  5569.618 &  -0.540 &   3.417 &   37.5 &   31.1 & \\
     Fe I  &  5572.842 &  -0.275 &   3.397 &   50.8 &   44.6 & \\
     Fe I  &  5576.088 &  -1.000 &   3.430 &   22.4 &   18.3 & \\
     Fe I  &  5586.755 &  -0.096 &   3.368 &   61.3 &   57.8 & \\
     Fe I  &  5615.644 &   0.050 &   3.332 &   72.2 &   67.6 & \\
     Fe I  &  5624.542 &  -0.755 &   3.417 &   26.9 &   22.5 & \\
     Fe I  &  5658.816 &  -0.793 &   3.397 &   26.8 &   22.6 & \\
     Fe I  &  5686.529 &  -0.450 &   4.549 &    3.0 &    ... & \\
     Fe I  &  5701.543 &  -2.216 &   2.559 &   14.8 &   12.2 & \\
     Fe I  &  6065.481 &  -1.530 &   2.609 &   43.3 &   39.4 & \\
     Fe I  &  6136.614 &  -1.400 &   2.453 &   66.1 &   58.3 & \\
     Fe I  &  6137.691 &  -1.403 &   2.588 &   56.5 &   48.4 & \\
     Fe I  &  6191.558 &  -1.420 &   2.433 &   64.0 &   56.9 & \\
     Fe I  &  6200.312 &  -2.437 &   2.609 &    9.4 &    8.1 & \\
     Fe I  &  6213.429 &  -2.480 &   2.223 &   20.7 &   16.7 & \\
     Fe I  &  6219.280 &  -2.433 &   2.198 &   26.8 &   23.6 & \\
     Fe I  &  6230.723 &  -1.281 &   2.559 &   67.9 &   60.8 & \\
     Fe I  &  6246.318 &  -0.733 &   3.603 &    ... &   16.0 & \\
     Fe I  &  6252.555 &  -1.687 &   2.404 &   53.8 &   48.6 & \\
     Fe I  &  6254.257 &  -2.443 &   2.279 &   24.4 &   21.7 & \\
     Fe I  &  6265.134 &  -2.550 &   2.176 &   24.9 &   20.1 & \\
     Fe I  &  6301.499 &  -0.718 &   3.654 &   17.5 &   14.2 & \\
     Fe I  &  6322.685 &  -2.426 &   2.588 &   11.2 &    7.8 & \\
     Fe I  &  6335.330 &  -2.180 &   2.198 &   37.7 &   34.9 & \\
     Fe I  &  6336.835 &  -1.050 &   3.686 &   12.9 &   10.5 & \\
     Fe I  &  6393.601 &  -1.432 &   2.433 &   61.0 &   53.2 & \\
     Fe I  &  6400.000 &  -0.290 &   3.603 &   39.5 &   33.6 & \\
     Fe I  &  6411.649 &  -0.595 &   3.654 &   23.1 &   20.0 & \\
     Fe I  &  6421.350 &  -2.027 &   2.279 &   46.3 &   39.6 & \\
     Fe I  &  6430.845 &  -2.006 &   2.176 &   55.6 &   50.2 & \\
     Fe I  &  6498.940 &  -4.699 &   0.958 &    9.7 &    ... & \\
     Fe I  &  6592.913 &  -1.470 &   2.728 &    ... &   33.2 & \\
     Fe I  &  6593.868 &  -2.422 &   2.433 &   15.9 &   13.7 & \\
     Fe I  &  6663.440 &  -2.479 &   2.424 &   15.9 &   12.9 & \\
     Fe I  &  6677.986 &  -1.420 &   2.692 &   50.7 &   44.5 & \\
     Fe I  &  6750.151 &  -2.621 &   2.424 &   12.2 &   14.0 & \\
     Co I  &  3842.047 &  -0.770 &   0.923 &   56.7 &   54.7 & \\
     Co I  &  3845.468 &   0.010 &   0.923 &   93.6 &   86.9 & \\
     Co I  &  3873.120 &  -0.660 &   0.432 &  129.9 &  121.4 & \\
     Co I  &  3881.869 &  -1.130 &   0.582 &   78.0 &   75.8 & \\
     Co I  &  3995.306 &  -0.220 &   0.923 &   82.7 &    ... & \\
     Co I  &  4020.898 &  -2.070 &   0.432 &   40.6 &   34.2 & \\
     Co I  &  4110.532 &  -1.080 &   1.049 &   39.0 &   31.7 & \\
     Co I  &  4121.318 &  -0.320 &   0.923 &   91.4 &   88.6 & \\
     Ni I  &  3807.144 &  -1.220 &   0.423 &  115.4 &  107.4 & \\
     Ni I  &  3858.301 &  -0.951 &   0.423 &  126.6 &  114.3 & \\
     Ni I  &  4648.659 &  -0.160 &   3.420 &   16.2 &   11.2 & \\
     Ni I  &  4714.421 &   0.230 &   3.380 &   31.6 &   22.8 & \\
     Ni I  &  4855.414 &   0.000 &   3.542 &    ... &   10.0 & \\
     Ni I  &  4980.161 &  -0.110 &   3.606 &   15.6 &   10.3 & \\
     Ni I  &  5017.591 &  -0.080 &   3.539 &   14.5 &    8.3 & \\
     Ni I  &  5035.374 &   0.290 &   3.635 &   21.7 &   16.4 & \\
     Ni I  &  5080.523 &   0.130 &   3.655 &   21.8 &   17.1 & \\
     Ni I  &  5084.081 &   0.030 &   3.678 &   12.4 &    7.3 & \\
     Ni I  &  5137.075 &  -1.990 &   1.676 &   35.5 &   26.1 & \\
     Ni I  &  5578.734 &  -2.640 &   1.676 &    6.4 &    4.1 & \\
     Ni I  &  5754.675 &  -2.330 &   1.935 &   11.8 &    9.6 & \\
     Ni I  &  6108.121 &  -2.450 &   1.676 &   10.9 &    7.3 & \\
     Ni I  &  6643.641 &  -2.300 &   1.676 &   25.1 &   19.6 & \\
     Ni I  &  6767.778 &  -2.170 &   1.826 &   23.4 &   14.5 & \\
     Zn I  &  4722.150 &  -0.390 &   4.030 &   14.2 &   11.1 & \\
     Zn II &  4810.530 &  -0.170 &   4.080 &   18.8 &   16.4 & \\
     Sc II &  4246.820 &   0.240 &   0.315 &  133.0 &  152.4 & \\
     Sc II &  4324.998 &  -0.440 &   0.595 &   96.7 &  109.7 & \\
     Sc II &  4415.544 &  -0.670 &   0.595 &   74.2 &   89.6 & \\
     Sc II &  5031.010 &  -0.400 &   1.357 &   41.9 &   63.9 & \\
     Sc II &  5239.811 &  -0.770 &   1.455 &   20.3 &   34.3 & \\
     Sc II &  5318.374 &  -2.010 &   1.357 &    2.9 &    6.3 & \\
     Sc II &  5526.785 &   0.020 &   1.768 &   41.4 &   63.8 & \\
     Sc II &  5641.000 &  -1.130 &   1.500 &   13.3 &   26.4 & \\
     Sc II &  5657.907 &  -0.600 &   1.507 &   32.1 &   52.0 & \\
     Sc II &  5658.362 &  -1.210 &   1.497 &   10.6 &   21.1 & \\
     Sc II &  5667.164 &  -1.310 &   1.500 &    8.3 &   19.3 & \\
     Sc II &  5669.055 &  -1.200 &   1.500 &   10.1 &   21.4 & \\
     Sc II &  5684.214 &  -1.070 &   1.507 &   11.8 &   25.6 & \\
     Sc II &  6604.578 &  -1.310 &   1.357 &    8.7 &   20.9 & \\
     Ti II &  3813.394 &  -2.020 &   0.607 &   97.0 &    ... & \\
     Ti II &  3882.291 &  -1.710 &   1.116 &    ... &   79.3 & \\
     Ti II &  4012.396 &  -1.750 &   0.574 &  104.8 &  113.5 & \\
     Ti II &  4025.120 &  -1.980 &   0.607 &   86.1 &   93.7 & \\
     Ti II &  4028.355 &  -1.000 &   1.892 &   55.4 &   68.7 & \\
     Ti II &  4053.829 &  -1.210 &   1.893 &   42.6 &   52.7 & \\
     Ti II &  4161.527 &  -2.160 &   1.084 &   53.3 &   63.1 & \\
     Ti II &  4163.634 &  -0.400 &   2.590 &   45.0 &   55.6 & \\
     Ti II &  4184.309 &  -2.510 &   1.080 &   30.7 &   42.8 & \\
     Ti II &  4330.723 &  -2.060 &   1.180 &   46.9 &   63.1 & \\
     Ti II &  4337.876 &  -1.130 &   1.080 &  101.4 &  112.9 & \\
     Ti II &  4395.833 &  -1.970 &   1.243 &   49.4 &   57.5 & \\
     Ti II &  4399.786 &  -1.270 &   1.237 &   97.1 &  102.2 & \\
     Ti II &  4417.715 &  -1.430 &   1.165 &   99.6 &  107.2 & \\
     Ti II &  4418.306 &  -1.990 &   1.237 &   53.2 &   60.0 & \\
     Ti II &  4441.731 &  -2.410 &   1.180 &   39.7 &   56.0 & \\
     Ti II &  4443.775 &  -0.700 &   1.080 &  119.9 &  128.3 & \\
     Ti II &  4444.536 &  -2.210 &   1.116 &   48.1 &   60.7 & \\
     Ti II &  4450.503 &  -1.510 &   1.084 &   88.5 &   92.1 & \\
     Ti II &  4464.461 &  -2.080 &   1.161 &   68.0 &   79.3 & \\
     Ti II &  4468.517 &  -0.600 &   1.131 &  121.7 &  132.5 & \\
     Ti II &  4470.835 &  -2.280 &   1.165 &   48.3 &   59.7 & \\
     Ti II &  4488.342 &  -0.820 &   3.124 &    7.7 &   16.1 & \\
     Ti II &  4501.269 &  -0.760 &   1.116 &  117.4 &  121.1 & \\
     Ti II &  4529.480 &  -2.030 &   1.572 &   44.0 &   56.7 & \\
     Ti II &  4533.972 &  -0.770 &   1.237 &  122.3 &  130.9 & \\
     Ti II &  4545.144 &  -1.810 &   1.131 &   32.3 &   44.8 & \\
     Ti II &  4563.770 &  -0.960 &   1.221 &  109.9 &  121.0 & \\
     Ti II &  4571.957 &  -0.530 &   1.572 &  110.8 &  123.6 & \\
     Ti II &  4583.408 &  -2.870 &   1.165 &   17.7 &   23.9 & \\
     Ti II &  4589.915 &  -1.790 &   1.237 &   73.9 &   86.2 & \\
     Ti II &  4657.212 &  -2.320 &   1.243 &   37.3 &   49.6 & \\
     Ti II &  4708.651 &  -2.370 &   1.237 &   33.8 &   47.7 & \\
     Ti II &  4779.979 &  -1.370 &   2.048 &   32.4 &   44.0 & \\
     Ti II &  4798.507 &  -2.670 &   1.080 &   31.9 &   43.7 & \\
     Ti II &  4805.089 &  -1.100 &   2.061 &   46.2 &   56.7 & \\
     Ti II &  4865.610 &  -2.810 &   1.116 &   25.6 &   35.1 & \\
     Ti II &  4911.175 &  -0.340 &   3.124 &    8.1 &   14.0 & \\
     Ti II &  5005.159 &  -2.730 &   1.566 &    7.9 &   12.6 & \\
     Ti II &  5129.155 &  -1.390 &   1.892 &   43.9 &   57.1 & \\
     Ti II &  5185.902 &  -1.350 &   1.893 &   36.7 &   50.3 & \\
     Ti II &  5188.691 &  -1.210 &   1.582 &   91.1 &  100.1 & \\
     Ti II &  5268.607 &  -1.620 &   2.598 &    3.8 &    9.4 & \\
     Ti II &  5336.778 &  -1.630 &   1.582 &   53.2 &   64.5 & \\
      V II &  3916.411 &  -1.053 &   1.428 &   34.6 &   45.6 & \\
      V II &  3951.960 &  -0.784 &   1.476 &   40.8 &   47.7 & \\
      V II &  4005.710 &  -0.522 &   1.820 &   32.4 &   42.4 & \\
      V II &  4023.378 &  -0.689 &   1.805 &   32.1 &   45.3 & \\
      V II &  4035.622 &  -0.767 &   1.793 &   48.9 &   52.4 & \\
     Fe II &  4178.855 &  -2.480 &   2.583 &   56.7 &   47.2 & \\
     Fe II &  4233.167 &  -2.000 &   2.583 &    ... &   77.2 & \\
     Fe II &  4416.817 &  -2.600 &   2.778 &   46.6 &   34.7 & \\
     Fe II &  4489.185 &  -2.970 &   2.828 &   24.1 &   21.8 & \\
     Fe II &  4491.401 &  -2.700 &   2.856 &   33.1 &   25.1 & \\
     Fe II &  4508.283 &  -2.580 &   2.856 &   51.5 &   43.6 & \\
     Fe II &  4515.337 &  -2.480 &   2.844 &   45.0 &   35.7 & \\
     Fe II &  4541.523 &  -3.050 &   2.856 &   22.0 &   15.5 & \\
     Fe II &  4555.890 &  -2.290 &   2.828 &   54.1 &   43.8 & \\
     Fe II &  4576.331 &  -2.940 &   2.844 &   22.1 &   18.2 & \\
     Fe II &  4583.829 &  -2.020 &   2.807 &   83.4 &   72.7 & \\
     Fe II &  4731.439 &  -3.360 &   2.891 &   17.7 &   10.2 & \\
     Fe II &  4923.930 &  -1.320 &   2.891 &  104.7 &   93.2 & \\
     Fe II &  4993.350 &  -3.670 &   2.810 &    7.3 &    ... & \\
     Fe II &  5018.450 &  -1.220 &   2.891 &  115.1 &  106.6 & \\
     Fe II &  5197.559 &  -2.100 &   3.230 &   38.7 &   30.1 & \\
     Fe II &  5234.619 &  -2.270 &   3.221 &   43.6 &   33.7 & \\
     Fe II &  5275.999 &  -1.940 &   3.199 &   50.8 &   40.2 & \\
     Fe II &  5284.080 &  -3.190 &   2.891 &   17.1 &   13.9 & \\
     Fe II &  5534.834 &  -2.930 &   3.245 &   15.4 &   12.9 & \\
     Fe II &  6247.545 &  -2.510 &   3.892 &    8.6 &    4.7 & \\
     Sr II &  4077.714 &   0.150 &   0.000 &  162.6 &  106.4 & \\
     Sr II &  4215.524 &  -0.180 &   0.000 &  151.1 &   95.3 & \\
     Ba II &  4554.033 &   0.163 &   0.000 &  100.3 &   58.0 & \\
     Ba II &  4934.086 &  -0.160 &   0.000 &   91.4 &   51.2 & \\
     Ba II &  6496.896 &  -0.369 &   0.604 &    ... &   12.1 & \\

\enddata
\end{deluxetable}

\begin{deluxetable}{lcccc}
\tablewidth{0pt}
\tablecaption{ATMOSPHERIC PARAMETERS \label{tab:param}}
\tablehead{
Star    & {\teff} &  {\logg} & [Fe/H] & {\vt} 
}
\startdata
{\BS} & 4500 & 1.0 & $-2.8$ & 2.1 \\ 
{\HD} & 4600 & 1.1 & $-2.6$ & 2.2 
\enddata
\end{deluxetable}

\begin{deluxetable}{cccccccccccc}
\tablewidth{0pt}
\tablecaption{ABUNDANCE RESULTS\label{tab:res}}
\tablehead{
Element & Species  &  Solar Abundance\tablenotemark{a} & \multicolumn{4}{c}{\BS}           & & \multicolumn{4}{c}{\HD} \\
\cline{4-7}\cline{9-12}
        &          & {\loge}          & {\loge} & [X/Fe] & $N$ & $\sigma$ & &  {\loge} & [X/Fe] & $N$ & $\sigma$ 
}
\startdata
C       & CH       &      8.39    & 5.35    & $-$0.27 & \nodata & 0.17 & & 5.45 & $-$0.35 & \nodata & 0.17 \\ 
N       & CN       &      7.78    & 5.65    &   +0.65 & \nodata & 0.29 & & 5.80 &   +0.61 & \nodata & 0.29 \\ 
O       & \ion{O}{1} &    8.66    & 7.03    &   +1.14  & 2    &  0.19     & & 6.81 &   +0.74 & 2       &  0.19\\ 
Na      & \ion{Na}{1} &   6.17    & 4.20    &   +0.81  & 2    &  0.13     & & 3.39 & $-$0.19 & 2       &  0.13\\ 
Mg      & \ion{Mg}{1} &   7.53    & 5.98    &   +1.23  & 6    &  0.13     & & 5.44 &   +0.49 & 6       &  0.11\\ 
Al      & \ion{Al}{1} &   6.37    & 3.76    &   +0.17  & 1    &  0.19     & & 3.38 & $-$0.40 & 1       &  0.18\\ 
Si      & \ion{Si}{1} &   7.51    & 5.17    &   +0.44  & 4    &  0.14     & & 5.34 &   +0.41 & 4       &  0.16\\ 
Ca      & \ion{Ca}{1} &   6.31    & 3.89    &   +0.35  & 22   &  0.13     & & 3.93 &   +0.20 & 20      &  0.11\\ 
Sc      & \ion{Sc}{2} &   3.05    & 0.87    &   +0.59  & 14   &  0.24     & & 0.57 &   +0.11 & 14      &  0.23\\ 
Ti      & \ion{Ti}{1} &   4.90    & 2.61    &   +0.48  & 28   &  0.13     & & 2.42 &   +0.11 & 27      &  0.12\\ 
Ti      & \ion{Ti}{2} &   4.90    & 2.74    &   +0.62  & 42   &  0.25     & & 2.58 &   +0.27 & 41      &  0.24\\ 
Cr      & \ion{Cr}{1} &   5.64    & 2.67    & $-$0.19  & 11   &  0.11     & & 2.57 & $-$0.38 & 13      &  0.10\\ 
Mn      & \ion{Mn}{1} &   5.39    & 2.01    & $-$0.61  & 6    &  0.12     & & 2.23 & $-$0.57 & 8       &  0.13\\ 
Fe      & \ion{Fe}{1} &   7.45    & 4.68    & $-$2.78 \tablenotemark{b} & 130  &  0.20     & & 4.86 & $-$2.59 \tablenotemark{b}& 214     &  0.20 \\
Fe      & \ion{Fe}{2} &   7.45    & 4.67    & $-$2.79 \tablenotemark{b} & 20   &  0.19     & & 4.87 & $-$2.58 \tablenotemark{b}& 20      &  0.19\\ 
Co      & \ion{Co}{1} &   4.92    & 2.34    &   +0.19  & 6    &  0.12     & & 2.48 &   +0.15 & 7       &  0.11\\ 
Ni      & \ion{Ni}{1} &   6.23    & 3.38    & $-$0.08  & 16   &  0.12     & & 3.66 &   +0.02 & 15      &  0.11\\ 
Zn      & \ion{Zn}{1} &   4.60    & 1.94    &   +0.11  & 2    &  0.22     & & 2.08 &   +0.06 & 2       &  0.21\\ 
Sr      & \ion{Sr}{2} &   2.92    &$-$1.22  & $-$1.36  & 2    &  0.25     & & 0.08 & $-$0.30 & 2       &  0.15\\ 
Ba      & \ion{Ba}{2} &   2.17    &$-$2.29  & $-$1.69  & 2    &  0.24     & &$-$1.69  & $-$1.28 & 2    &  0.23
\enddata
\tablenotetext{a}{\citet{asplund05}}
\tablenotetext{b}{[Fe/H] values}
\end{deluxetable}

\clearpage

\end{document}